\documentclass[final,3p,times,twocolumn]{elsarticle}

\usepackage{epsfig}

\usepackage{amssymb}
\usepackage{amsmath}

\usepackage{color}
\usepackage{float}

\journal{Nuclear Physics B Proceedings Supplement}

\begin{document}

\begin{frontmatter}



\title{Weak annihilation rare radiative B-decays}


\author[msu,sinp]{Anastasiia Kozachuk}
\author[msu,sinp,itep]{Nikolai Nikitin}

\address[msu]{M.~V.~Lomonosov Moscow State University, Physical Facutly, 119991, Moscow, Russia}
\address[sinp]{D.~V.~Skobeltsyn Institute of Nuclear Physics, M.~V.~Lomonosov
Moscow State University, 119991, Moscow, Russia}
\address[itep]{A.~I.~Alikhanov Institute for Theoretical and Experimental Physics, 117218 Moscow, Russia}

\begin{abstract}
We give predictions for the branching ratios of several rare radiative B-decays which proceed only through weak annihilation mechanism. In addition to previous analysis we estimate the branching ratios of $\bar{B}^0_s\to D^{*0}\gamma$ and $B_c^-\to\rho^-\gamma$ decays and obtain $2.87\cdot 10^{-9}$ and $2.59\cdot 10^{-7}$ for them correspondingly. The accuracy of the predictions is at the level of 20\%. 
\end{abstract}


\end{frontmatter}


\section{Introduction}
The investigation of rare $B$ decays forbidden at the tree level in the Standard Model provides the 
possibility to probe the electroweak sector at large mass scales. Interesting information 
about the structure of the theory is contained in the Wilson coefficients entering the 
effective Hamiltonian which take different values in different theories with testable 
consequences in rare $B$ decays. 

There is an interesting class of rare radiative $B$-decays which proceed merely through the weak annihilation mechanism. 
These processes have very small probabilities and have not been observed. 
So far, only upper limits on the branching ratios of these decays have been obtained: In 2004, the BaBar Collaboration 
provided the upper limit ${\cal B} (B^0\to J/\psi\gamma)<1.6\cdot 10^{-6}$ \cite{BaBar2004}. 
Last year, the LHCb Collaboration reached the same sensitivity to the $B^0$-decay and set the limit on the $B_s^0$ decay: 
${\cal B} (B^0\to J/\psi\gamma)<1.7\cdot 10^{-6}$ and ${\cal B} (B_s^0\to J/\psi\gamma)<7.4\cdot 10^{-6}$ at 90\% CL 
\cite{LHCb2015}. 
Obviously, with the increasing statistics, the prospects to improve the limits on the branching ratios by one order of magnitude 
or eventually to observe these decays in the near future seem very favorable. 

In this paper we consider more annihilation type decays in addition to our recent work \cite{Koza}. The paper is organized as follows: In Section \ref{Sect2} the effective weak Hamiltonian and the structure of the amplitude are recalled.   
In Section \ref{Sect3} we consider the photon emission from the $B$-loop and present the $B\to\gamma$ transition 
form factors within the 
relativistic dispersion approach based on constituent quark picture. 
Section \ref{Sect4} contains the analysis of the $V\to\gamma$ transition form factors. 
Finally, in Section \ref{Sect5} the numerical estimates are given. 

\section{\label{Sect2}The effective Hamiltonian, the amplitude, and the decay rate}
We consider the weak annihilation radiative $B\to V\gamma$ transition, where 
$V$ is the vector meson. The corresponding amplitude is given by the matrix element of the effective Hamiltonian \cite{heff}
\begin{eqnarray}
A(B\to V\gamma)=\langle \gamma(q_1)V(q_2)|H_{\rm eff}|B(p) \rangle,  
\end{eqnarray}
where $p$ is the $B$ momentum, $q_2$ is the vector-meson momentum, and $q_1$ is the photon
momentum, $p=q_1+q_2$, $q_1^2=0$, $q_2^2=M_V^2$, $p^2=M_B^2$. 
The effective weak Hamiltonian of the transition has the form (we provide in this 
Section formulas for the effective Hamiltonian with the flavor structure 
$\bar d c\, \bar u b$, but all other decays of interest may be easily described by an obvious replacement of the 
quark flavors and the corresponding CKM factors $\xi_{\rm CKM}$):
\begin{eqnarray}
\label{Heff}
H_{\rm eff} = 
-\frac{G_F}{\sqrt{2}}{\xi_{\rm CKM}}
\Big(C_1(\mu){\cal O}_1+C_2(\mu){\cal O}_2\Big), 
\end{eqnarray}
$G_F$ is the Fermi constant, $\xi_{\rm CKM}=V^*_{cd}V_{ub}$, $C_{1,2}(\mu)$ are the scale-dependent Wilson coefficients \cite{heff}, 
and we only show the relevant four-quark operators 
\begin{eqnarray}
{\cal O}_1 = \bar d_{\alpha}\gamma_{\nu}(1-\gamma_5)c_{\alpha}\;
\bar u_{\beta}\gamma_{\nu}(1-\gamma_5) b_{\beta},\nonumber
\\
{\cal O}_2 = \bar d_{\alpha}\gamma_{\nu}(1-\gamma_5) c_{\beta}\; 
\bar u_{\beta}\gamma_{\nu}(1-\gamma_5) b_{\alpha}.
\end{eqnarray}
We use notations $e=\sqrt{4\pi\alpha_{\rm em}}$, 
$\gamma^5=i\gamma^0\gamma^1\gamma^2\gamma^3$,  
$\sigma_{\mu\nu}=i\left [\gamma_{\mu},\gamma_{\nu}\right ]/2$,
$\epsilon^{0123}=-1$ and 
${\rm Sp}\left (\gamma^5\gamma^{\mu}\gamma^{\nu}\gamma^{\alpha}\gamma^{\beta}\right )
=4i\epsilon^{\mu\nu\alpha\beta}$.
It is convenient to parameterize the amplitude in the following way  
\begin{equation}
\begin{split}
\label{F_PC}
A=\frac{eG_F}{\sqrt{2}}
\Big[\epsilon_{q_1\epsilon^\ast_1 q_2 \epsilon_2^\ast}F_{\rm PC}\\
+i \epsilon_2^{\ast\nu}\epsilon_1^{\ast\mu} \left(g_{\nu\mu}\,pq_1-p_\mu q_{1\nu}\right)F_{\rm PV}
\Big], 
\end{split}
\end{equation}
where $F_{\rm PC}$ and $F_{\rm PV}$ are the parity-conserving and 
parity-violating invariant amplitudes, respectively. Here $\epsilon_2$($\epsilon_1$) is the 
vector-meson (photon) polarization vector. We use the short-hand notation 
$\epsilon_{abcd}=\epsilon_{\alpha\beta\mu\nu}a^{\alpha}b^{\beta}c^{\mu}d^{\nu}$ 
for any 4-vectors $a,b,c,d$. 

For the decay rate one finds 
\begin{equation}
\begin{split}
\label{rate}
\Gamma(B\to V\gamma)=\frac{G^2_F\,\alpha_{em}}{16}M_B^3
\left(1-{M^2_V}/{M_B^2}\right)^3 \\
     \times\left( |F_{\rm PC}|^2+|F_{\rm PV}|^2 \right). 
\end{split}
\end{equation}
Within the naive factorization approach the matrix element of the amplitude can be simplified and the parity-conserving and parity-violating parts of the amplitude can be represented as follows \cite{Koza}:
\begin{eqnarray}
\label{fpc}
F_{\rm PC}=\xi_{\rm CKM} a_{\rm eff}(\mu)\left[\frac{F_V}{M_B}f_VM_V +f_B H_P\right],
\end{eqnarray}
\begin{equation}
\begin{split}
F_{\rm PV}=\xi_{\rm CKM}a_{\rm eff}(\mu)\Big[\frac{F_{A}}{M_B}f_V M_V+f_B H_S \\
-\frac{2Q_Bf_Bf_VM_V}{M_B^2-M_V^2}\Big].
\end{split}
\end{equation}  

Summing up this Section, within the factorization approximation the weak annihilation amplitude may be expressed in terms of four form factors: 
$F_A$, $F_V$, $H_P$ and $H_S$. 
It should be emphasized that each of the form factors $F_A$, $F_V$, $H_P$ and $H_S$ actually depends on two variables: 
The $B$-meson transition form factors $F_A$, $F_V$ depend on $q_1^2$ and $q_2^2$, and  
$F_{A,V}(q_1^2,q_2^2)$ should be evaluated at $q_1^2=0$ and $q_2^2=M_V^2$. 
The vector-meson transition form factors $H_P$ and $H_S$ 
depend on $q_1^2$ and $p^2$, and $H_{S,P}(q_1^2,p^2)$ should be evaluated at $q_1^2=0$ and $p^2=M_B^2$.

\section{\label{Sect3}Photon emission from the $B$-meson loop and the form factors $F_A$ and $F_V$.}
In this section we calculate the form factors $F_{A,V}$ within the relativistic quark model, which is a dispersion approach based on constituent quark picture. This approach has been formulated in detail in \cite{melikhov} and applied to the weak decays of heavy mesons in \cite{ms}.

The pseudoscalar meson in the initial state is described in the dispersion approach by the following vertex \cite{m}: 
$\bar q_1(k_1)\; i\gamma_5 q(-k_2)\;G(s)/{\sqrt{N_c}}$, 
with $G(s)=\phi_P(s)(s-M_P^2)$, $s=(k_1+k_2)^2$, $k_1^2=m_1^2$ and $k_2^2=m_2^2$. 
We represent the pseudoscalar-meson wave function $\phi_P(s)$ in the form $\phi_P(s)=f(s)w(k^2)$, $k^2=\lambda(s,m_1^2,m^2)/4s$. For $w(k^2)$ we use a simple Gaussian parametrization $w(k^2)=A(\beta)e^{-k^2/(2\beta^2)}$. For the concrete form of $f(s)$ and the normalization condition which gives $A(\beta)$ see \cite{m}. The parameter $\beta$ is extracted from the condition
\begin{eqnarray}
\label{fP}
f_P=\sqrt{N_c}\int\limits_{(m_1+m_2)^2}^\infty ds \phi_P(s)\rho_{P}(s), 
\end{eqnarray}
with
\begin{eqnarray}
\begin{split}
\rho_{P}(s)=(m_1+m_2)\frac{\lambda^{1/2}(s,m_1^2,m_2^2)}{8\pi^2s}\\\times\frac{s-(m_1-m_2)^2}{s}.
\end{split}
\end{eqnarray}
Here $\lambda(a,b,c)=(a-b-c)^2-4bc$ is the triangle function.  

\begin{figure}[H]
\mbox{\epsfig{file=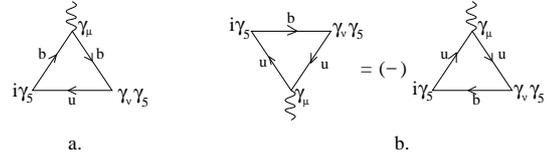,width=7cm}}
\caption{\label{fig:Fa}Diagrams for the form factor $F_A$: a) $F_A^{(b)}$, b) $F_A^{(u)}$.} 
\end{figure}

\begin{figure}[H]
\mbox{\epsfig{file=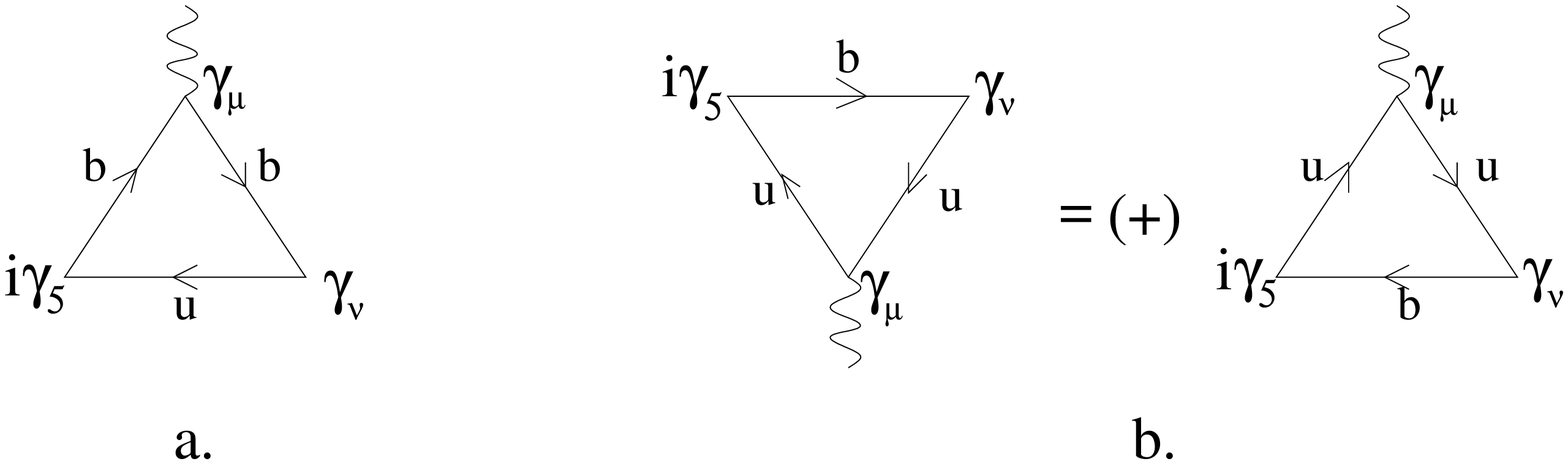,width=7cm}}
\caption{\label{fig:Fv}Diagrams for the form factor $F_V$: a) $F_V^{(b)}$, b) $F_V^{(u)}$.} 
\end{figure}

Recall that the form factors $F_{A,V}$ describe the transition of the $B$-meson to the photon with the momentum 
$q_1$, $q_1^2=0$, induced by the axial-vector (vector) current with the momentum $q_2$, $q_2^2=M_V^2$.
The form factors $F_{A,V}$ are given by the diagrams of Fig \ref{fig:Fa} and \ref{fig:Fv}. 

\subsection{The form factor $F_A$}
Let's consider form factor $F_{A}$ first.
Fig \ref{fig:Fa}a shows $F_A^{(b)}$, the contribution to the form factor of the process when the $b$ quark interacts with the 
photon; Fig \ref{fig:Fa}b describes the contribution of the process when the quark $u$ interacts while 
$b$ remains a spectator. It is convenient to change the direction of the quark line in the loop diagram of Fig \ref{fig:Fa}b. This is done by performing the charge conjugation of the matrix element and leads to a sign change for the $\gamma_\nu\gamma_5$ vertex. For the form factor we then obtain the following expression:    
\begin{eqnarray}
F_A=Q_b F_A^{(b)}-Q_u F_A^{(u)},
\end{eqnarray} 
where $F_A^{(b)}=F_{A}^{(1)}(m_b,m_u)$, $F_A^{(u)}=F_{A}^{(1)}(m_u,m_b)$, and
\begin{equation}
\begin{split}
\label{fadisp}
\frac{1}{M_B}F_{A}^{(1)}(m_1,m_2)=\frac{\sqrt{N_c}}{4\pi^2}\int\limits_{(m_1+m_2)^2}^\infty
\frac{ds\;\phi_B(s)}{(s-M_V^2)}\\
\times\left(\rho_+(s,m_1,m_2)+2\frac{m_1-m_2}{s-M_V^2}\rho_{k_\perp^2}(s,m_1,m_2)\right), 
\end{split}
\end{equation}
with
\begin{equation}\label{rhoplus}
\begin{split}
\rho_+(s,m_1,m_2)=(m_2-m_1)\frac{\lambda^{1/2}(s,m_1^2,m_2^2)}{s}\\
+m_1\log\left(\frac{s+m_1^2-m_2^2+\lambda^{1/2}(s,m_1^2,m_2^2)}{s+m_1^2-m_2^2-\lambda^{1/2}(s,m_1^2,m_2^2)}\right),
\end{split}
\end{equation}
\begin{equation}\label{rhokperp2}
\begin{split}
\rho_{k_\perp^2}(s,m_1,m_2)=\frac{s+m_1^2-m_2^2}{2s}\lambda^{1/2}(s,m_1^2,m_2^2)\\-
m_1^2\log\left(\frac{s+m_1^2-m_2^2+\lambda^{1/2}(s,m_1^2,m_2^2)}
{s+m_1^2-m_2^2-\lambda^{1/2}(s,m_1^2,m_2^2)}\right).
\end{split}
\end{equation}
Employing the fact that the wave function $\phi_B(s)$ is localized near the threshold in the region $\sqrt{s}-m_b-m_u\le \bar\Lambda$, it is easy to show that in the limit $m_b\to\infty$ the photon emission from the light quark dominates over the emission from the heavy quark \cite{korch}
\begin{eqnarray}
\label{Fa_HQ}
\frac{1}{M_B}F_{A}^{(u)}=\frac{f_B}{\bar\Lambda m_b}+...,\qquad  \frac{1}{M_B}F_{A}^{(b)}=\frac{f_B}{m_b^2}+...
\end{eqnarray}

\subsection{The form factor $F_V$}
The consideration of the form factor $F_{V}$ is very similar to the form factor $F_{A}$. 
$F_V$ is determined by the two diagrams shown in Fig \ref{fig:Fv}:  
Fig \ref{fig:Fv}a gives $F_V^{(b)}$, the contribution of the process when the $b$ quark interacts with the photon; Fig \ref{fig:Fv}b 
describes the contribution of the process when the quark $u$ interacts. 

It is again convenient to change the direction of the quark line in the loop diagram of Fig \ref{fig:Fv}b by performing the charge 
conjugation of the matrix element. For the vector current $\gamma_\nu$ in the vertex the sign does not change (in contrast to the $\gamma_\nu\gamma_5$ case considered above). The calculation of $F_V$ within the dispersion approach gives the following result:  
\begin{eqnarray}
F_V=Q_bF_V^{(b)}+Q_uF_V^{(u)}, 
\end{eqnarray}
where $F_V^{(b)}=F_V^{(1)}(m_b,m_u)$, $F_V^{(u)}=F_V^{(1)}(m_u,m_b)$, and
\begin{equation}
\begin{split}
\frac{1}{M_B}F_V^{(1)}(m_1,m_2)=\\-\frac{\sqrt{N_c}}{4\pi^2}\int\limits_{(m_1+m_2)^2}^\infty
\frac{ds\phi_B(s)}{(s-M_V^2)}\rho_+(s,m_1,m_2). 
\end{split}
\end{equation}
The function $\rho_+(s,m_1,m_2)$ is given in (\ref{rhoplus}). 
In the heavy-quark limit $m_b\to\infty$ one finds 
\begin{eqnarray}
\label{Fv_HQ}
\frac{1}{M_B}F_{V}^{(u)}=-\frac{f_B}{\bar\Lambda m_b}+..., \\
\nonumber
\qquad \frac{1}{M_B}F_{V}^{(b)}=-\frac{f_B}{m_b^2}+...
\end{eqnarray}
The dominant contribution in the heavy quark limit again comes from the process 
when the light quark emits the photon. As seen from Eqs. (\ref{Fa_HQ}) and (\ref{Fv_HQ}), 
one finds $F_A=F_V$ in the heavy quark limit, in agreement with the large-energy effective theory \cite{leet}. 

\section{\label{Sect4}Photon emission from the vector meson loop. The form factors $H_S$ and $H_P$.}
We now calculate the form factors $H_{P,S}$ using the relativistic quark model. The vector meson in the final state is described in this approach by the vertex $\bar q_2(-k_2)\Gamma_\beta q_1(k_1')$, 
$\Gamma_\beta=\left(-\gamma_\beta+\frac{(k_1'-k_2)_\beta}{\sqrt{s}+m_1+m_2}\right)\;G(s)/{\sqrt{N_c}}$, 
with $G(s)=\phi_V(s)(s-M_V^2)$, $s=(k'_1+k_2)^2$, ${k'}_1^2=m_1^2$ and $k_2^2=m_2^2$. The vector meson wave function $\phi_V(s)$ has the same structure and normalizing conditions as that of the pseudoscalar meson described in Section \ref{Sect3}. All the details were given in \cite{m}. For obtaining the parameter $\beta$ we use the following condition
\begin{eqnarray}
\label{fV}
f_V=\sqrt{N_c}\int\limits_{(m_1+m_2)^2}^\infty ds \phi_V(s)\rho_{V}(s),
\end{eqnarray}
with
\begin{equation}
\begin{split}
\rho_{V}(s)=\frac{2\sqrt{s}+m_1+m_2}{3}\frac{\lambda^{1/2}(s,m_1^2,m_2^2)}{8\pi^2s} \\ \times\frac{s-(m_1-m_2)^2}{s}.
\end{split}
\end{equation}
 
\begin{figure}[H]
\mbox{\epsfig{file=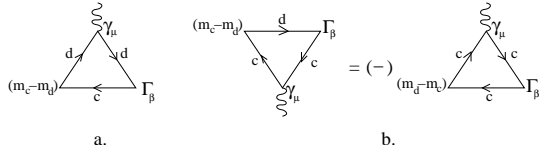,width=7cm}} 
\caption{\label{fig:Hs}Diagrams for the form factor $H_S$: a) $H_S^{(d)}$, b) $H_S^{(c)}$.} 
\end{figure}

\begin{figure}[H]
\mbox{\epsfig{file=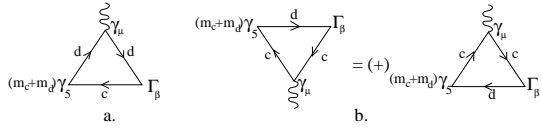,width=7cm}}
\caption{\label{fig:Hp}Diagrams for the form factor $H_P$: a) $H_P^{(d)}$, b) $H_P^{(u)}$.} 
\end{figure}

The form factors $H_S$ and $H_P$ are given by the diagrams of Fig \ref{fig:Hs} and \ref{fig:Hp}.

\subsection{The form factor $H_S$}
Fig \ref{fig:Fa}a shows $H_S^{(d)}$, the contribution to the form factor of the process when the $d$ quark interacts with the 
photon; Fig \ref{fig:Hs}b describes the contribution of the process when the quark $u$ interacts while 
$d$ remains spectator. 
Changing the direction of the quark line in the loop diagram of 
Fig \ref{fig:Hs}b, leads to a sign change for the scalar current $j=(m_c-m_d)\bar d c$ in the vertex, and we derive the following formula:       
\begin{eqnarray}
H_S=Q_d H_S^{(d)}-Q_c H_S^{(c)},
\end{eqnarray} 
with $H_S^{(d)}=H_S^{(1)}(m_d,m_c)$, $H_S^{(c)}=H_S^{(1)}(m_c,m_d)$,
\begin{equation} \label{hsdisp}
\begin{split}
H_S^{(1)}(m_1,m_2)=\frac{\sqrt{N_c}}{4\pi^2}\int\limits_{(m_1+m_2)^2}^\infty
\frac{ds\;\phi_V(s)}{(s-p^2-i0)}(m_2-m_1)\\ \times
\left(\rho_+(s,m_1,-m_2)+\frac{2\sqrt{s}}{p^2-M_V^2}\rho_{k_\perp^2}(s,m_1,m_2)\right), 
\end{split}
\end{equation}
where $\rho_+(s,m_1,m_2)$ and $\rho_{k_\perp^2}(s,m_1,m_2)$ are determined earlier in (\ref{rhoplus}) and (\ref{rhokperp2}).

One can obtain the behaviour of the form factor in the limit $m_Q\to\infty$ 
for the heavy-light vector meson $\bar Qq$ (and assuming $p^2\sim m_Q^2$): 
\begin{eqnarray}
\label{Hs_HQ}
H_S^{(q)}\propto f_V/\bar \Lambda,\qquad H_S^{(Q)}\propto f_V/m_Q,
\end{eqnarray}
but one expects a strong numerical suppression because of the partial cancellation of the leading-order contributions. 

\subsection{The form factor $H_P$}
The form factor $H_{P}$ is determined by the two diagrams shown in Fig \ref{fig:Hp}:  
Fig \ref{fig:Hp}a gives $H_P^{(d)}$, the contribution of the process when the $d$-quark interacts with the 
photon; Fig \ref{fig:Hp}b describes the contribution of the process when the $c$-quark interacts. 
We again change the direction of the quark line in the loop diagram of 
Fig \ref{fig:Hp}b by performing the charge conjugation of the matrix element. 
For the pseudoscalar current $(m_c+m_d)\bar d\gamma_5 c$ in the vertex the sign does not change and we obtain:
\begin{eqnarray}
H_P=Q_dH_P^{(d)}+Q_cH_P^{(c)}, 
\end{eqnarray}
where $H_P^{(d)}=H_P^{(1)}(m_d,m_c)$, $H_P^{(c)}=H_P^{(1)}(m_c,m_d)$, 
\begin{equation}
\begin{split}
H_P^{(1)}(m_1,m_2)=\frac{\sqrt{N_c}}{4\pi^2}\int\limits_{(m_1+m_2)^2}^\infty
\frac{ds\phi_V(s)}{(s-p^2-i0)} \\ \times
(m_1+m_2)\left(\rho_+(s,m_1,m_2)+\frac{\rho_{k_\perp^2}}{\sqrt{s}+m_1+m_2}\right),  
\end{split}
\end{equation}
with $\rho_+(s,m_1,m_2)$ and $\rho_{k_\perp^2}(s,m_1,m_2)$ given in (\ref{rhoplus}) and (\ref{rhokperp2}). 
For the behaviour of $H^{(Q,q)}_P$ at large-$m_Q$ for the heavy-light vector 
meson $\bar Qq$ we get 
\begin{eqnarray}
\label{Hp_HQ}
H_P^{(q)}\to \frac{f_V}{\bar \Lambda}\frac{m_Q^2}{m_Q^2-p^2},\qquad H_P^{(Q)}\to \frac{f_V m_Q}{m_Q^2-p^2}. 
\end{eqnarray}

For the $B$-decays of interest, we need the value of the form factors $H_{P,S}(p^2,q_1^2=0)$ at $p^2=M_B^2$,
which lies above the threshold $(m_c+m_q)^2$. The spectral representations for $H_{P,S}(p^2=M_B^2)$ develop the imaginary parts 
which occur due to the quark-antiquark intermediate states in the $p^2$-channel. It should be emphasized that no anomalous cuts 
emerge if one consider the form factors in the double spectral representation at $q_1^2\le 0$ \cite{lms_triangle}. In all cases considered in this paper, the value of $p^2=M_B^2$ lies far above the region of resonances which occur in the quark-antiquark channel.  
Far above the resonance region, local quark-hadron duality works well and the calculation of the imaginary part based on the 
quark diagrams is trustable. The imaginary part turns out to be orders of magnitude smaller than the real part of the form factor 
and for the practical purpose of the decay rate calculation may be safely neglected. 

\section{\label{Sect5}Numerical results}
We have received the spectral representations of the form factors and now can obtain numerical estimations for the form factors and, finally, for the branching ratios. 
\subsection{Parameters of the model}
We use here the following values of the constituent quark masses:
\begin{equation}\label{quark_masses}
\begin{split}
\quad m_d=m_u=0.23 \;{\rm GeV}, \quad m_s=0.35 \;{\rm GeV}, \\ 
\quad m_c=1.45 \;{\rm GeV},\quad m_b=4.85 \;{\rm GeV}.
\end{split}
\end{equation} 
With the quark masses (\ref{quark_masses}) and the meson wave function parameters $\beta$ quoted 
in Table \ref{table:parameters}, the decay constants from the dispersion 
approach reproduce the well-known decay constants of pseudoscalar and vector mesons also summarized in Table 
\ref{table:parameters}. 
\begin{table}[H]
\caption{\label{table:parameters}
Meson masses from \cite{pdg}, leptonic decay constants, and the corresponding wave function parameters $\beta$ \cite{lmss}.}
\centering
\begin{tabular}{|c|c|c|c|}
\hline
        & $M$, GeV  &  $f$, MeV &  $\beta$, GeV \\
\hline
$B$     & 5.279     & $192\pm 8$ \cite{lmsfB}    & $0.565$ \\
\hline
$B_s$   & 5.370     & $226\pm 15$ \cite{lmsfB}   & $0.62$ \\
\hline
$B_c$   & 6.275     & $427\pm 6$  \cite{hpqcd12}  & $0.94$ \\
\hline
$D^*$   & 2.010     & $248\pm 2.5$ \cite{lmsfD*} & $0.48$ \\
\hline
$D^*_s$     &  2.11     & $311\pm 9$ \cite{lmsfD*}       & $0.54$ \\
\hline
$J/\psi$    &  3.097    & $405\pm 7$ \cite{fpsi_lat,pdg} &  $0.68$ \\
\hline
$\rho$      &  0.775    & $209\pm 2$  \cite{pdg}  &   0.31  \\
\hline
\end{tabular}
\end{table}

\subsection{Predictions for the decay rates}
Using the values given above we now obtain the estimations for the branching ratios. 
We present results for six weak annihilation rare radiative B-decays:  
\begin{eqnarray}
\label{br1}
{\cal B}(\bar B^0_s\to J/\psi\gamma)    &=&1.2\cdot 10^{-7}\left(\frac{a_2}{0.15}\right)^2, \\
\label{br2}
{\cal B}(\bar B^0_d\to J/\psi\gamma)    &=&5.5\cdot 10^{-9}\left(\frac{a_2}{0.15}\right)^2, \\
\label{br3}
{\cal B}(\bar B^0_d\to D^{0*}\gamma)    &=&5.2\cdot 10^{-8}\left(\frac{a_2}{0.15}\right)^2,  \\
\label{br4}
{\cal B}(\bar B^0_s\to D^{0*}\gamma)    &=&2.9\cdot 10^{-9}\left(\frac{a_2}{0.15}\right)^2,  \\
\label{br5}
{\cal B}(B^-\to D_s^{*-}\gamma)    &=&2.1\cdot 10^{-7}\left(\frac{a_1}{1.02}\right)^2, \\
\label{br6}
{\cal B}(B^-_c\to \rho^-\gamma)    &=&2.6\cdot 10^{-7}\left(\frac{a_1}{1.02}\right)^2.
\end{eqnarray}
For the scale-dependent Wilson coefficients $C_i(\mu)$ and $a_{1,2}(\mu)$ at the renormalization scale $\mu\simeq 5$ GeV 
we use the following values \cite{heff}: $C_1=1.1$, $C_2=-0.241$, $a_1=C_1+C_2/N_c=1.02$, and $a_2=C_2+C_1/N_c=0.15$. 

\section*{\label{Sect6}Conclusions}
We obtained predictions for the branching ratios of weak annihilation rare radiative B-decays. The calculations were done in the naive factorization approximation, taking into account both the photon emission from the $B$-meson loop and the vector-meson loop ($V$-loop). The form factors containing non-perturbative QCD effects were calculated within the relativistic quark model. In comparison to previous analysis we obtained estimations for the decays $\bar{B}^0_s\to D^{*0}\gamma$ and $B_c^-\to\rho^-\gamma$. The accuracy of the predictions is at the level of 20\%. 

\section*{Acknowledgments} 
The authors are thankful to Dmitri Melikhov for useful discussions. The work was supported by grant 16-12-10280 of the Russian Science
Foundation. 





\end{document}